\documentclass[aps,prd,amsmath,amssymb,twocolumn,groupedaddress]{revtex4-2}  
\usepackage{graphicx}
\usepackage{hyperref}
\usepackage{natbib}
\usepackage{xcolor}

\definecolor{dbl}{rgb}{0, 0, 0.9}
\hypersetup{
  bookmarksopen=true,     
  colorlinks=true,        
  linkcolor=dbl,     
  citecolor=dbl,      
  filecolor=dbl,      
  urlcolor=dbl        
}

\begin{document}

\title{UHECR deflections in the coherent Galactic magnetic field}
\author{Alexander Korochkin}
\email{alexander.korochkin@ulb.be}
\affiliation{
 Université Libre de Bruxelles, CP225 Boulevard du Triomphe, 1050 Brussels, Belgium
}
\author{Dmitri Semikoz}
\affiliation{
 Université de Paris Cite, CNRS, Astroparticule et Cosmologie, F-75013 Paris, France
}
\author{Peter Tinyakov}
\affiliation{
 Université Libre de Bruxelles, CP225 Boulevard du Triomphe, 1050 Brussels, Belgium
}
\date{\today}

\begin{abstract}
We study the deflections of ultra-high-energy cosmic rays in several widely used models of the coherent Galactic magnetic field (GMF), including \texttt{PT11} (Pshirkov et al.~\cite{Pshirkov:2011um}), \texttt{JF12} (Jansson and Farrar~\cite{JF_GMF_1}), \texttt{UF23} (Unger and Farrar~\cite{Unger:2023lob}) and \texttt{KST24} (Korochkin, Semikoz, and Tinyakov~\cite{Korochkin:2024yit}). We propagate particles with rigidities of $5$, $10$, and $20$~EV and analyze the differences in deflection predictions across these GMF models. We identify the GMF components responsible for deflections in various regions of the sky and discuss the uncertainties in modeling these components, as well as potential future improvements.
\end{abstract}

\maketitle

\section{Introduction}

Two modern ultra-high-energy cosmic ray (UHECR) experiments, the Pierre Auger Observatory (Auger)~\cite{PierreAuger:2015eyc} and the Telescope Array (TA)~\cite{Tokuno:2012mi,TelescopeArray:2012uws}, were largely motivated by the possibility of UHECR astronomy. At that time, it was believed that most of the UHECR above the ankle at $E>3$~EeV were protons \cite{Berezinsky:2002nc}. In this case, the deflections of UHECR with $E>60$~EeV in the Galactic magnetic field (GMF) with the magnitude of $B_\mathrm{Gal} \sim \mu$G were expected to be on the order of a few degrees \cite{Tinyakov:2001ir}. The extragalactic field was not expected to produce large deflections either. Indeed, since protons lose energy in interactions with the cosmic microwave background (CMB) radiation \cite{Greisen:1966jv,Zatsepin:1966jv}, their propagation distance is limited to $50-100$~Mpc. Taking into account constraints on the intergalactic magnetic field (IGMF) $B_\mathrm{IGMF} \lesssim 1$~nG
the latter would not significantly change the trajectories of UHECR protons on their way from sources to our Galaxy. As a result, most UHECR protons with $E>60$~EeV should point back to their sources within a few degrees, and one could try to identify the sources after collecting a few years of data with the Auger and TA experiments. Moreover, exact modeling of the GMF would not be required if one could identify the sets of events of different energies coming from the same source \cite{Golup:2009cv,Giacinti:2009fy}. 

However, the actual situation appears to be more challenging for UHECR astronomy. In 2009, the Auger Collaboration presented evidence, based on the analysis of the distribution of air shower maxima, indicating that the composition of cosmic rays shifts from intermediate to heavy nuclei at the highest energies \cite{PierreAuger:2010ymv}. For heavy or even intermediate nuclei, the role of the GMF model is significantly amplified, as is the effect of its uncertainties. For instance, in the case of iron nuclei, deflections in the GMF are expected to be of order $60-90^\circ$  even for $E>60$ EeV  UHECR \cite{Giacinti:2010dk,Giacinti:2010ep}, leading to uncertainties in reconstructed source positions that may reach several tens of degrees.
This makes search for sources of UHECR extremely difficult. 

The observed UHECR flux exhibits a remarkable degree of isotropy. The most significant deviation ---in fact, the only one to surpass the $5\sigma$ significance threshold--- is the dipole anisotropy detected by Auger at $E>8$~EeV \cite{PierreAuger:2017pzq, PierreAuger:2024fgl}. A combined analysis of data from Auger and TA has yielded an all-sky measurement of the flux multipoles \cite{TelescopeArray:2021ygq, PierreAuger:2023mvf} in an assumption-free manner. Additionally, several intermediate-scale anisotropies have been observed. Earlier experiments had already provided some evidence of such anisotropies \cite{Kachelriess:2005uf}. Currently, in the southern sky, Auger detects the excess of events in the direction of the Cen A radio galaxy \cite{PierreAuger:2010ofq, PierreAuger:2024hrj}, while in the northern sky, TA observes a hotspots \cite{TelescopeArray:2014tsd}. The nature of these excesses, however, is not yet firmly established. 

The observed isotropy of UHECR suggests that these particles undergo significant deflections in magnetic fields. This makes it nearly impossible to identify their sources by directly cross-correlating observed UHECR directions with catalogs of potential sources, except in cases where the catalogs contain only a few objects. An alternative approach is to reconstruct the source directions by backtracking the observed events through the GMF, assuming the field follows one of the existing models \cite{Pshirkov:2011um, JF_GMF_1, Han2018, Shaw:2022lqd, Unger:2023lob, Han2024, Korochkin:2024yit}. 
Likewise, applying correction for GMF one can attempt to explain the dipole and intermediate angular scale anisotropies in the arrival directions of UHECR \cite{Allard:2021ioh, Allard:2023uuk, Bister:2023icg, Bister:2024ocm}. Backtracking is particularly promising for the highest-energy events, such as the Amaterasu particle, which was observed by TA and had an energy of $E\sim 244$~EeV \cite{TelescopeArray:2023sbd}. 

To carry out these tasks, precise knowledge of GMF is essential. In this context, it is crucial to identify regions of the sky where GMF is reasonably well understood and the predicted deflections can be trusted,  at least as a first approximation. These are, in the first place, the regions where the deflections are minimal. 

In this paper, we compare the deflections of UHECR predicted by our recent GMF model \cite{Korochkin:2024yit} with those expected in alternative or earlier models, such as \cite{Pshirkov:2011um}, \cite{JF_GMF_1}, and \cite{Unger:2023lob}. Through this comparison, we identify the components of each model that contribute to the deflections in different regions of the sky and determine the areas where these deflections are likely to be minimal.

Since the composition of UHECR is uncertain and likely changes with energy 
\cite{PierreAuger:2010ymv,PierreAuger:2022atd,PierreAuger:2023htc,PierreAuger:2024hlp}, we present our results for UHECR deflections in terms of rigidity $R=E/Z$ rather than energy. 
At $E=10$~EeV protons have the rigidity of $R=10$~EV. On the other hand, the highest energy events with energies in excess of $100$~EeV, assuming most of them are heavy nuclei, also have rigidities around $10$~EV. To account for potential uncertainties,
we present the comparison of deflections at three values of rigidity, $R=5$~EV, $R=10$~EV, and $R= 20$~EV. These values likely represent the rigidity range of the bulk of UHECR.

In Section \ref{sec:defl}, we compare the UHECR deflections predicted by different GMF models and discuss the sources of their uncertainties. This is followed by a summary and concluding remarks in Section \ref{sec:summary}.

\section{GMF model dependence of UHECR deflections}
\label{sec:defl}
The GMF is an unavoidable medium that deflects charged UHECRs, thereby obscuring their sources. While the general structure of the GMF is consistent across recent models, the parametrization and exact profiles of its components differ, making the UHECR deflection highly model-dependent. In this section, we compare the predicted UHECR deflections in popular coherent GMF models, including the two most recent models, \texttt{UF23} and \texttt{KST24}. We discuss the GMF components responsible for UHECR deflections in certain regions of the sky, as well as the modeling uncertainties associated with these components and possible future improvements. Our analysis is focused on the coherent GMF; however, deflections by the turbulent GMF are briefly discussed in Section~\ref{sec:defl_turb}.

\subsection{Deflection magnitudes}
\label{sec:defl_magn}
We start our comparison by reconstructing UHECR deflections in four models of the coherent GMF: \texttt{PT11} (Pshirkov et al.~\cite{Pshirkov:2011um}), \texttt{JF12} (Jansson and Farrar~\cite{JF_GMF_1}), \texttt{UF23~base} (Unger and Farrar~\cite{Unger:2023lob}) and \texttt{KST24} (Korochkin, Semikoz, and Tinyakov~\cite{Korochkin:2024yit}, the C++ implementation of \texttt{KST24} model can be found at \footnote{\url{https://zenodo.org/records/14743599}}). To obtain the directions of the particle momenta before entering the GMF, we backtracked antiparticles from the position of the Solar System to the edge of the Galaxy, defined as a galactocentric radius of $r=20$~kpc. As the propagation distances are small compared to typical UHECR mean free paths, interactions were neglected.

\begin{figure*}
    \centering
    \includegraphics[width=0.9\linewidth]{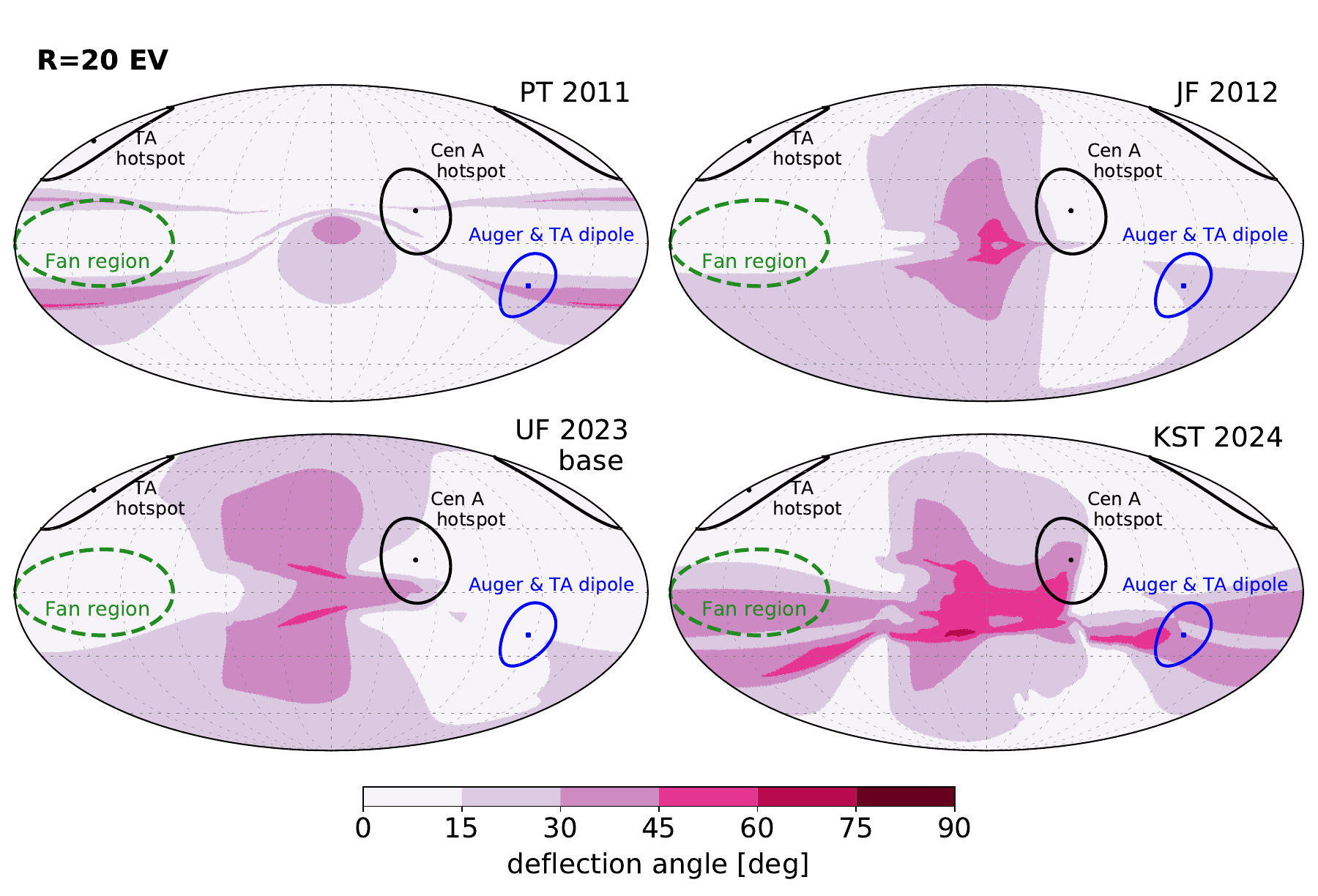}
    \includegraphics[width=0.9\linewidth]{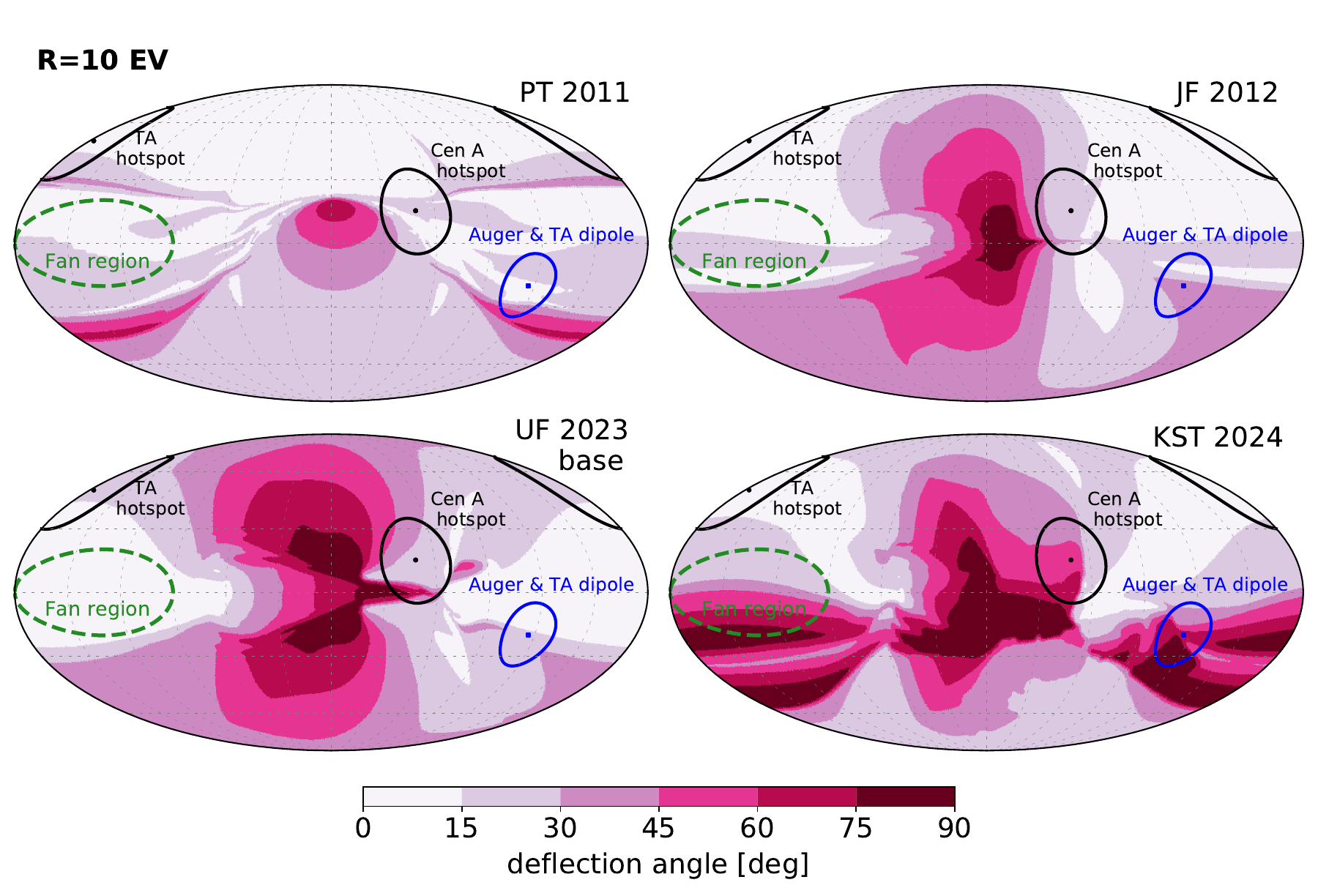}
    \caption{Angular deflections of $R=20$~EV (upper panels) and $R=10$~EV (lower panels) particles in \texttt{PT11}, \texttt{JF12}, \texttt{UF23~base} and \texttt{KST24} models of the coherent GMF. The position of the dipole is for the energies $E^\mathrm{TA}_\mathrm{Auger} \ge {}^{10~\mathrm{Eev}}_{8.55~\mathrm{Eev}}$ from~\cite{PierreAuger:2023mvf}. The skymaps are shown in the Galactic coordinates using the Mollweide projection. The Galactic center is at the center of the maps and longitude increases to the left. Note that all small-scale features visible on the maps will be washed out by the turbulent GMF, see Sec.~\ref{sec:defl_turb} for details.}
    \label{fig:defl_skymaps}
\end{figure*}

\begin{figure*}
    \centering
    \includegraphics[width=0.9\linewidth]{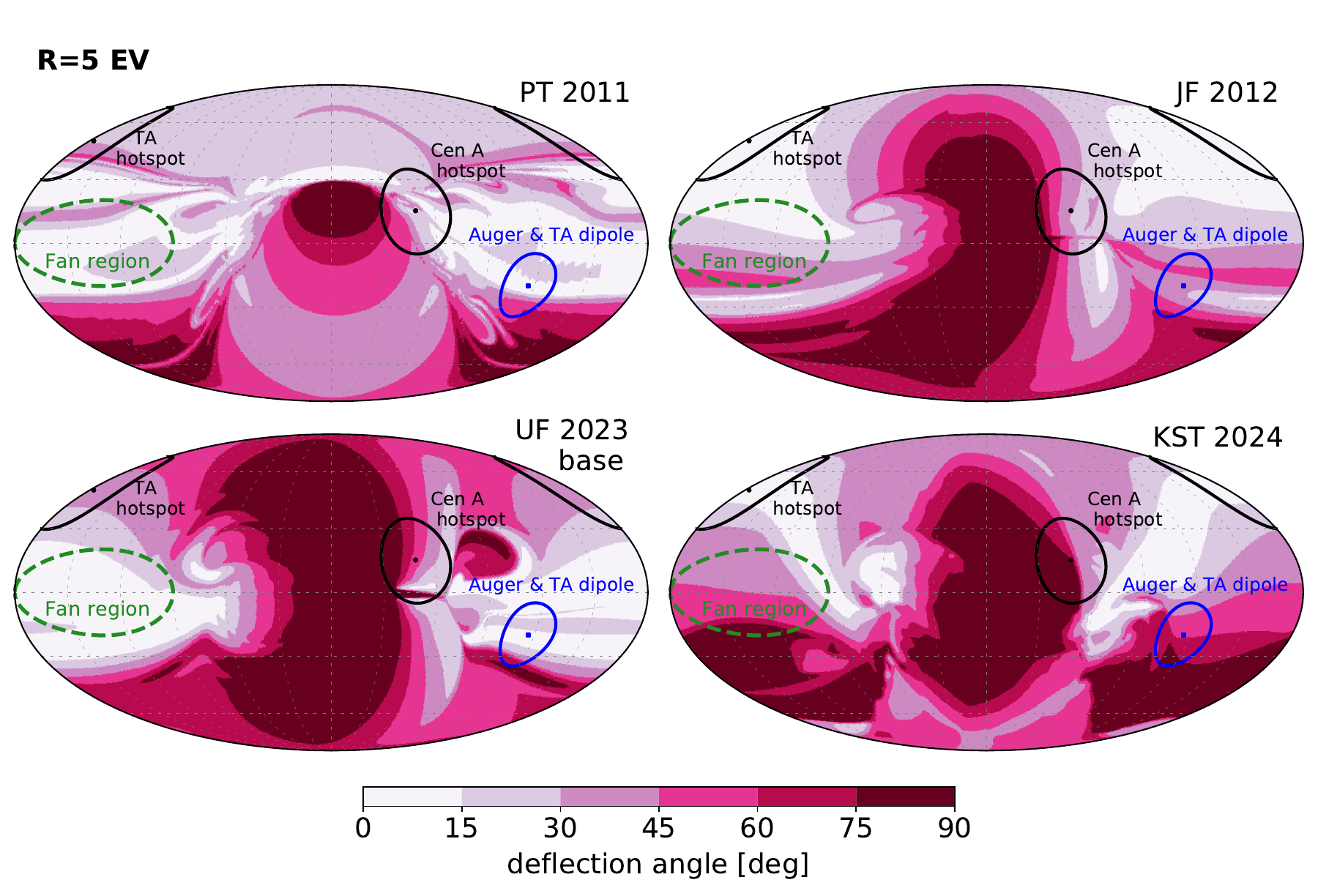}
    \caption{Same as Fig.~\ref{fig:defl_skymaps} but for particles with a rigidity of $R=5$~EV. }
    \label{fig:defl_skymaps5}
\end{figure*}

The results of the backtracking are shown in Fig.~\ref{fig:defl_skymaps} for particles with rigidities of $R=20$~EV and $R=10$~EV and in Fig.~\ref{fig:defl_skymaps5} for $R=5$~EV. Specifically, we measure the angular difference between the particles' momentum at the Solar System and at the edge of the Galaxy and plot it as a function of the former. The quantitative and qualitative differences between the different models are clearly visible. 

The total difference shown in Figs.~\ref{fig:defl_skymaps} and~\ref{fig:defl_skymaps5} accumulates effects of various origins. The first effect can be attributed to the different nature of the magnetic field tracers used to construct the GMF model. The most pronounced effect of this type is visible in the central part of the Galaxy ($-60^\circ < l < 60^\circ$). The \texttt{PT11} model predicts smaller deflections in this region of the sky compared to all other models. This is because \texttt{PT11} was fitted only to the rotation measures (RMs) of extragalactic sources, whereas three other models were also fitted to synchrotron data from the WMAP/\textit{Planck}. RM data is sensitive to the magnetic field parallel to the line of sight (LOS) and is thus almost insensitive to the X-field component of the GMF, which crosses the Galactic plane from below upwards in the central part of the Galaxy and is mainly perpendicular to the LOS. The presence of the X-field component is clearly identifiable in the synchrotron data and is primarily responsible for the large deflections observed in the central Galaxy at all latitudes in \texttt{JF12}, \texttt{UF23} and \texttt{KST24}.

The second type of difference is associated with uncertainties in the background fields and the different parameterizations of known GMF components. Indeed, typical GMF tracers are integral in nature and represent a combination of the magnetic field convolved with the thermal electron density (in the case of RM) or the cosmic-ray lepton density (in the case of synchrotron emission). Therefore, uncertainties in the background fields propagate into uncertainties in the GMF. Moreover, reconstructing the 3D magnetic field structure from 2D skymaps allows significant freedom in the parameterization of GMF components. 

Recently, a significant effort was made in \cite{Unger:2023lob} to explore this type of uncertainty. In this study, the authors considered several background electron density models along with a variety of GMF halo profiles to quantitatively estimate the resulting differences. As a result, a collection of eight significantly different models was presented. In Fig.\ref{fig:defl_skymaps}, we showed only the \texttt{base} model from the \texttt{UF23} collection. For UHECR deflections in the remaining seven \texttt{UF23} models and the related discussion, we refer the reader to~\cite{Unger:2023lob}, particularly to Figs.~18,~19~and~20 therein.

The last type of uncertainty is related to the interpretation of the data itself. Namely, the local features appearing in the data should be masked or modeled independently, while the large-scale features, associated with the coherent GMF, should be included in the model. The main obstacle here is that information about the distance to magnetic structures is still very limited and available only for a few of them. The interpretation of the data is one of the primary sources (along with parametrizations) of the differences between the most recent models, \texttt{UF23} and \texttt{KST24}. While both were fitted to very similar datasets (\texttt{UF23} and \texttt{KST24} were fitted to the latest compilation of the RMs of extragalactic sources~\cite{2023ApJS..267...28V} and polarized synchrotron data from WMAP/\textit{Planck} at 23/30~GHz~\cite{WMAP_23GHz}), the predicted UHECR deflections differ quite substantially in a significant fraction of the sky.

To better understand the differences between \texttt{UF23} and \texttt{KST24} models, it is useful to divide the sky into three regions: the central Galaxy ($-60^\circ < l < 60^\circ$), the outer Galaxy at high latitudes, and the outer Galaxy at low latitudes. In the central Galaxy the deflections are controlled by the X-field and its interplay with the toroidal field of the halo (the GMF halo component parallel to the disk). As shown in the Fig.~\ref{fig:defl_skymaps} the deflections in this region form a "butterfly-like" pattern. The typical deflections are large, exceeding $>20^\circ$ for $20$~EV particles even far above and below the Galactic plane. This is a robust prediction of the recent models \texttt{JF12}, \texttt{UF23}, and \texttt{KST24} all of which take into account the polarized synchrotron data. 

On the other hand, the boundaries of the "butterfly-like" region of strong deflections are not well constrained. As shown in Fig.~\ref{fig:defl_skymaps}, the shape of the "butterfly" is highly model-dependent. Moreover, the transition from large to small deflections occurs rapidly. This is particularly important because of the Cen A excess in the UHECR data, detected by Auger. This excess may indicate the presence of a local UHECR source and is located at the boundary of the "butterfly" region in all models, as shown by the black circle in Fig.~\ref{fig:defl_skymaps}. Technically, the instability in the Cen A region arises because the strongest X-field is concentrated in the inner 5-7 kpc around the Galactic center. Since GMF models are typically fit to the entire sky at once, the fit is not sensitive to the exact shape of the X-field profile at the edge of this region. However, this area is crucial for the UHECRs backtracking around the Cen A excess. A necessary improvement when constructing the next generation of GMF models should be a detailed analysis of this region.  

In contrast, in the outer Galaxy at high latitudes ($|b| \gtrsim 20^\circ$), all coherent field models predict relatively small deflections, likely not exceeding $30^\circ$. Thus, if the random deflections are similarly small (see Section~\ref{sec:defl_turb}), this region may be a promising place to search for UHECR sources. Interestingly, the TA hotspot, marked by the black circle in Fig.~\ref{fig:defl_skymaps}, is located in this region.

Finally, in the outer part of the sky at low Galactic latitudes ($90^\circ \gtrsim l \gtrsim 270^\circ$, $|b| \lesssim 30^\circ$), the deflections predicted by \texttt{KST24} differ drastically from those of \texttt{UF23}, see Fig.~\ref{fig:defl_skymaps}. The difference arises from the different treatment of the Fan region, an area of strong polarized synchrotron emission located in the outer Galaxy near the Galactic plane. In Fig.~\ref{fig:defl_skymaps} the Fan region is roughly shown with a green dashed contour. For a long time, the Fan region was believed to be a nearby magnetized bubble and was therefore excluded from GMF fitting. However, recent evidence~\cite{fan,2024arXiv240603765P} strongly suggests that the Fan region is a Galactic-scale feature that must be incorporated into coherent GMF models. The \texttt{KST24} model is the first to include the Fan region as part of the large-scale GMF.

Fitting the Fan region in \texttt{KST24} required a relatively strong magnetic field in the Local and Perseus Galactic spiral arms, which accounts for the stronger UHECR deflections predicted by \texttt{KST24} in the outer Galaxy compared to earlier models. However, there are two important caveats to consider. First, in the \texttt{KST24} analysis, the magnetic field strength in the Fan region is degenerate with the density of cosmic-ray leptons. As a result, the parameters of the magnetic field may change once a more reliable distribution of relativistic leptons becomes available. Second, a magnetic field reversal (from clockwise to counterclockwise) in the outer Galaxy could reduce the predicted deflections without affecting much the polarized synchrotron emission. While indication for such a reversal has been observed in pulsar data~\cite{Han2018}, its significance remains uncertain. In view of the large size of the Fan region, the magnitude of deflections in this region is essential for the  interpretation of the dipole observed by Auger \cite{PierreAuger:2017pzq,PierreAuger:2024fgl}.
\begin{figure}
    \centering
    \includegraphics[width=0.99\linewidth]{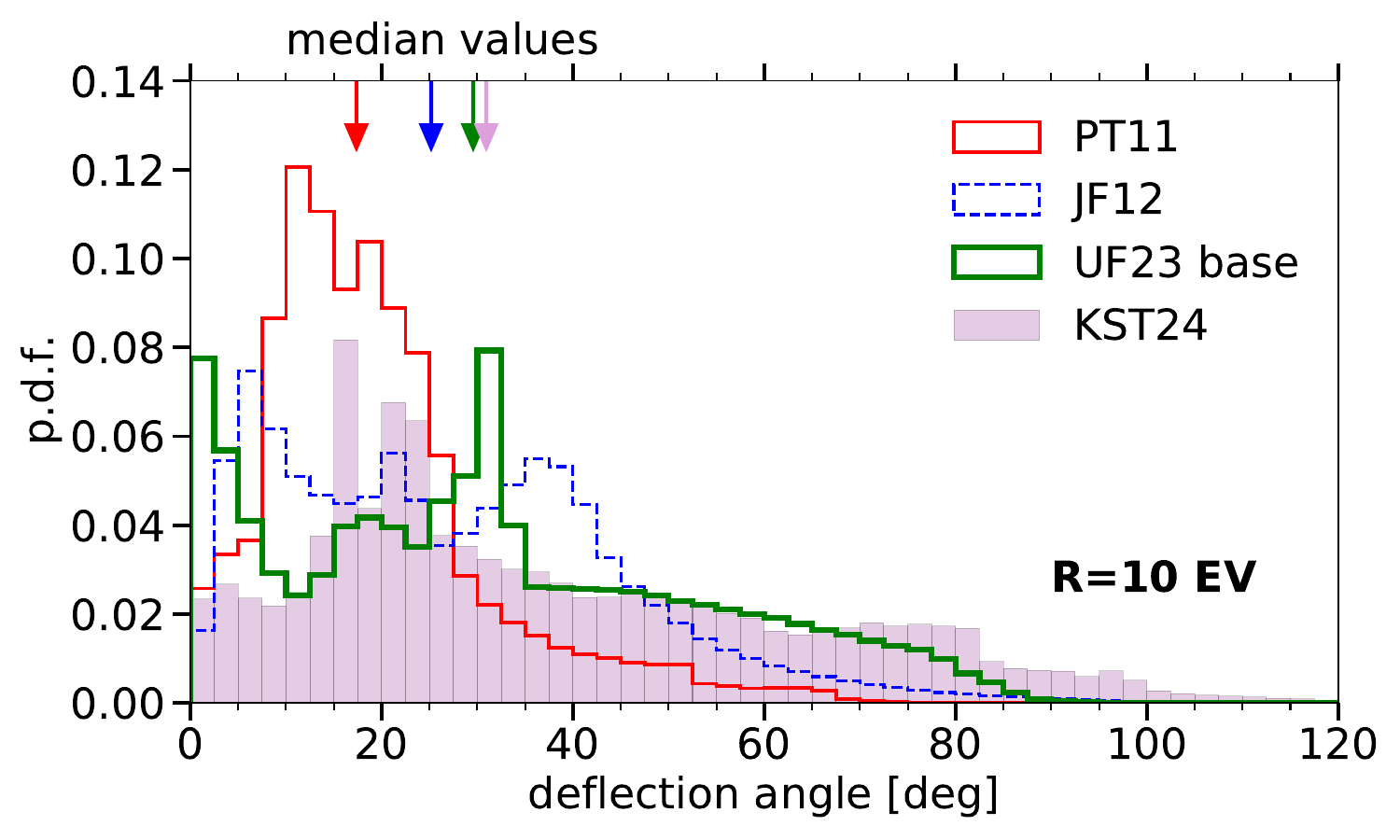}
    \caption{Histogram of angular deflections of $R=10$~EV\ particles, backtracked in \texttt{PT11}, \texttt{JF12}, \texttt{UF23 base} and \texttt{KST24} GMF models. The initial momenta of the particles were uniformly distributed over the sky, corresponding to the centers of the pixels in the HEALPix binning scheme with $\mathrm{NSIDE}=128$. The arrows indicate the medians of the corresponding distributions.}
    \label{fig:hist}
\end{figure}

Another new feature introduced in the \texttt{KST24} model is the magnetic field associated with the wall of the Local Bubble \cite{Korochkin:2024yit, LocalBubble_2}. The bubble wall itself does not strongly affect deflections due to its small size. The analysis by \cite{Korochkin:2024yit} demonstrated that it allows avoiding the need to invoke striated magnetic fields. Future studies will further clarify the role of the Local Bubble in UHECR deflections.

The all-sky distributions of deflection magnitudes in the four models are compared in Fig.~\ref{fig:hist} for $R=10$~EV particles. One can see that with the addition of GMF components the expected deflections increase. Median deflection in the \texttt{KST24} model of approximately~$31^\circ$, is nearly twice that of the \texttt{PT11} model, where it is about~$17^\circ$. Additionally, the \texttt{JF12} and \texttt{UF23 base} models, which do not fit the Fan~region, predict a significant portion of the sky with small deflections. This results in a peak of the corresponding histograms below $10^\circ$. As a consequence of fitting the Fan~region in the \texttt{KST24} model, these regions largely disappear, and the fraction of the sky with deflections smaller than $10^\circ$ drops to less than $\lesssim 10\%$.

\subsection{Deflection directions}
\label{sec:defl_dir}
\begin{figure*}
    \begin{minipage}[t]{0.483\textwidth}
        \includegraphics[scale=0.4]{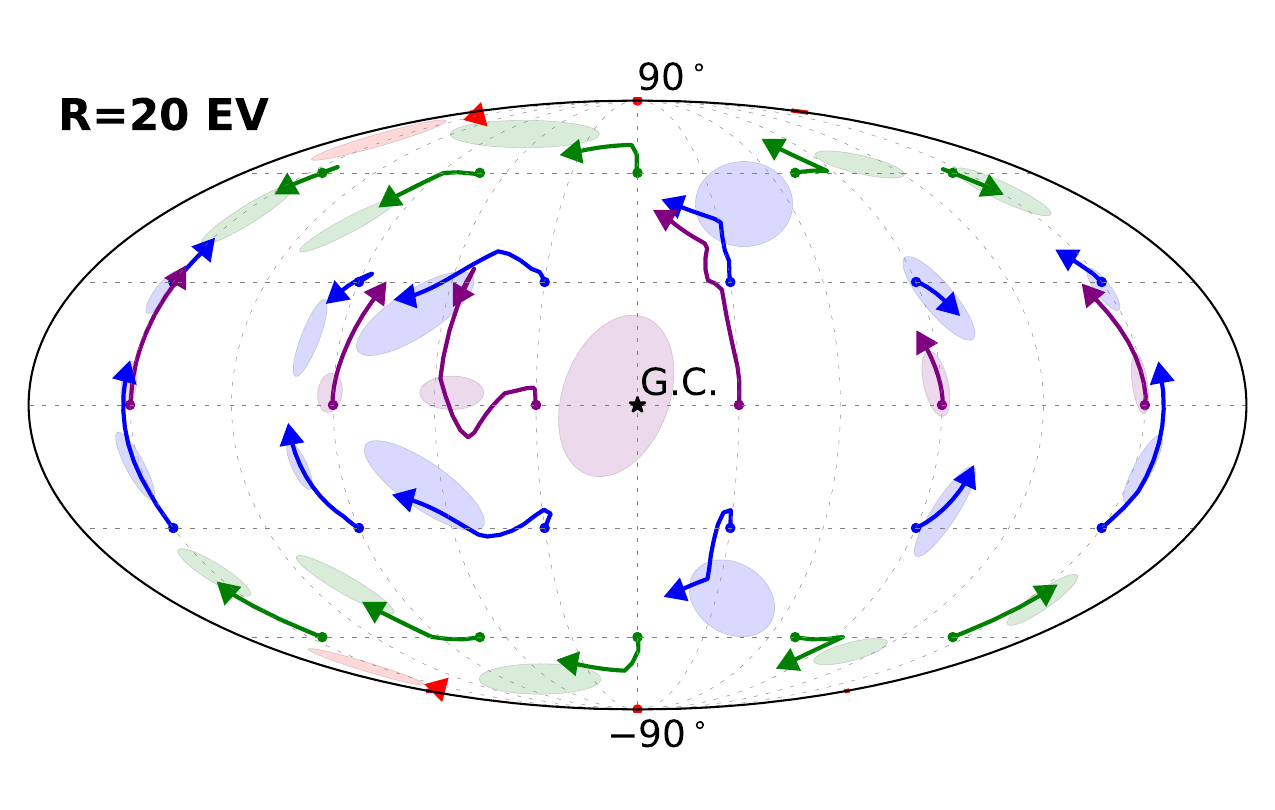}
        \caption{Directions of deflections of $R=20$~EV particles after backtracking in the \texttt{JF12}, \texttt{UF23} (all eight models), and \texttt{KST24} GMF models. The initial directions of the particles are shown with dots. The final directions at the edge of the Galaxy are marked with arrows for the \texttt{KST24} model and encircled with colored ellipses for the \texttt{JF12} and \texttt{UF23} collection of models. Different colors correspond to different particles' initial latitudes for better visibility. }
        \label{fig:defl_dir}
    \end{minipage}
    \hfill
    \begin{minipage}[t]{0.483\textwidth}
        \includegraphics[scale=0.4]{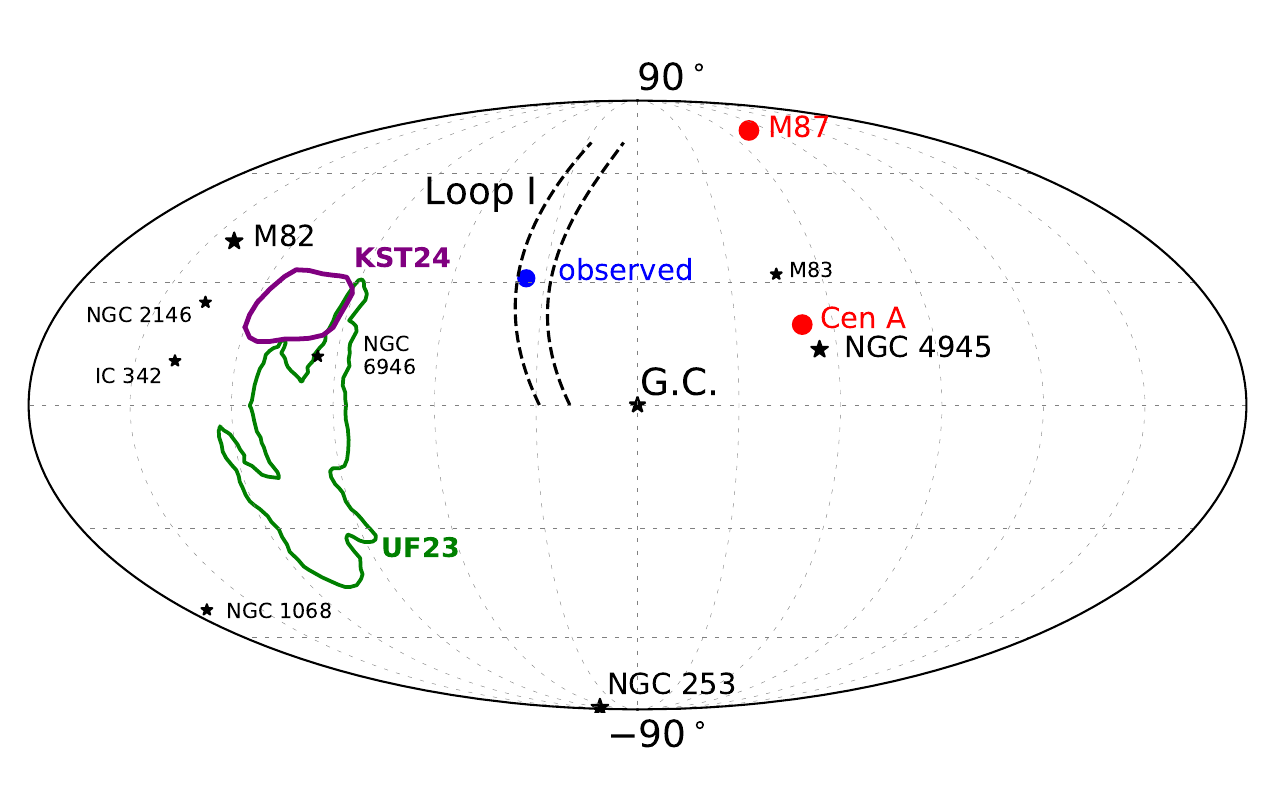}
        \caption{Localization of the source of the Amaterasu particle. The observed direction of Amaterasu is shown with a blue dot. A purple contour shows the localization using \texttt{KST24} model assuming Amaterasu is an iron nucleus. A green contour corresponds to a localization region obtained in \cite{Unger:2023hnu} for the \texttt{UF23} collection of models. For both contours the Amaterasu energy was rescaled to $E=212$~EeV to account for the systematic bias of the energy reconstruction for the heavy primaries~\cite{TelescopeArray:2023sbd}. Black stars mark bright starburst galaxies from the catalog of \cite{2019JCAP...10..073L}.}
        \label{fig:defl_amat}
    \end{minipage}
\end{figure*}

In this section, we continue the comparison and discuss the directions of UHECR deflections in the GMF models. The directions of deflections of $R=20$~EV particles are shown in Fig.~\ref{fig:defl_dir}. The particles were backtracked starting from a regular grid, marked with colored dots. The lines indicate the evolution of the direction of the particles' momentum as they were backtracked to the edge of the Galaxy in the \texttt{KST24} GMF model. The ellipses encircle the predictions of the \texttt{JF12} and \texttt{UF23} collection of models (including all eight models) for the same initial momenta of particles. One can see that, in roughly half of the sky, the predictions of the \texttt{KST24} GMF model lie outside the uncertainty regions of the \texttt{JF12} and \texttt{UF23} models. The largest differences are concentrated along the Galactic plane. This is to be expected, as this is the most difficult for modeling region of the Galaxy. 

As a particular example of the application of the \texttt{KST24} model, in Fig.~\ref{fig:defl_amat} we show the source localization region of the Amaterasu particle. This particle, detected by TA \cite{TelescopeArray:2023sbd}, had an arrival direction $(l,b)=(36.2^\circ,30.9^\circ)$ in Galactic coordinates and an extremely high energy of $E=244\pm29\mathrm{\,(stat.)^{+51}_{-76}\mathrm{\,(syst.)}}$~EeV. The origin of Amaterasu has been studied extensively \cite{2024JCAP...04..042K, Unger:2023hnu, Bourriche:2024bbe, 2025arXiv250106677M}, but its source has not yet been reliably identified.

In Fig.~\ref{fig:defl_amat}, we present the Amaterasu source localization. The purple contour for the \texttt{KST24} model was obtained by varying the model parameters, as well as particle energy, within their one-sigma uncertainty. Additionally, $5^\circ$ smearing by a toy model of the turbulent GMF was taken into account, following the prescription from~\cite{2013MNRAS.436.2326P}. For the \texttt{KST24} contour, the Amaterasu particle was assumed to be an iron nucleus. The contour size is primarily driven by the observed energy uncertainty. This, however, does not include ``systematic'' uncertainties related to modeling of GMF, so the actual uncertainties are larger. This is illustrated by the green contour in Fig.~\ref{fig:defl_amat}, which shows the source localization from a Monte Carlo analysis of  \cite{Unger:2023hnu} for the \texttt{UF23} models. Note that, for both the green contour from \cite{Unger:2023hnu} and the purple contour presented in this paper for \texttt{KST24}, the particles' energy was rescaled to $212$~EeV to account for systematic bias in energy reconstruction for heavy primaries. Note also that the observed direction of the Amaterasu event points close to the Loop~I, the bright feature in the polarized synchrotron data (shown with black dashed lines in~Fig.~\ref{fig:defl_amat}). Recently, it was shown to be a Galactic-scale outflow \cite{2024NatAs...8.1416Z, 2024A&A...691L..22C}, but it is not included in any GMF model. Incorporating this feature into GMF models could influence the reconstructed direction.

\subsection{Deflections in turbulent GMF}
\label{sec:defl_turb}
In the previous Sections, we discussed UHECR deflections in four different models of the coherent GMF. At the same time, the GMF also has a turbulent component, whose effect on UHECR deflections has been studied extensively in the literature (see, for instance~\cite{Harari:1999it, Harari:2002dy}.)

A general conclusion from these studies is that the coherent and turbulent components of the GMF affect UHECRs in qualitatively different ways: the coherent GMF shifts the entire source image, while the turbulent GMF produces a halo around the central position determined by the coherent field. Based on the analysis of the RM data,~\cite{Tinyakov:2004pw, 2013MNRAS.436.2326P} concluded that the typical deflection angle of particle with the rigidity of $R=10$~EV due to the turbulent GMF is of the order of 10$^\circ$ above the Galactic plane. A similar conclusion was reached by~\cite{Jansson:2012rt, 2019JCAP...05..004F} in their analysis of the total synchrotron intensity measured by WMAP. If so, the deflection due to turbulent GMF dominate only in a small fraction of the sky, in particular at high Galactic latitudes in the direction of the outer Galaxy, see Fig.~\ref{fig:defl_skymaps}.

In principle, the deflections due to turbulent field may be stronger if the Galaxy is embedded in a turbulent magnetized halo of $\sim 100$~kpc size filled with a strong magnetic field. There is an observational evidence from the low frequency radio observatory LOFAR, as well as the new X-ray observations showing the presence of $10^7$~K gas around the galaxies and in particular around our Galaxy~\cite{2024A&A...681A..78L, Blunier:2024aqx}. Such a halo could have been produced, for example, by stellar feedback in and around the Galactic disk~\cite{2023MNRAS.520.2655V,2024ApJ...975...49B}. At present, the level of magnetization and the turbulent spectrum of the halo remain unknown. A strongly magnetized halo, with a magnetic field strength exceeding $\sim 1$~$\mu$G and extending up to the virial radius around the Galaxy ($\sim 100$~kpc) can significantly affect UHECR trajectories. This possibility was considered in~\cite{Shaw:2022lqd, 2025arXiv250116881S}, although the existence of such a strongly magnetized halo has not yet been directly confirmed.

\section{Discussion and Summary}
\label{sec:summary}
GMF models have significantly improved from the previous generation models \texttt{PT11} and \texttt{JF12} to recent versions \texttt{UF23} and \texttt{KST24}. While the primary components of the GMF in the halo remain consistent with those in earlier models --- including the toroidal field below and above the Galactic plane and the X-shaped field in the central Galaxy --- the parametrization of these components and the results of their fits to observational data differ substantially. Notably, recent fits indicate generally higher magnitudes for the GMF, leading to increased typical expected deflections.

The situation is more challenging in the Galactic disk due to strong contamination from local structures, such as supernova bubbles, and the alignment of spiral arms along the line of sight, especially toward the Galactic center. The differences between models are most pronounced in the disk, making deflections at low Galactic latitudes the most uncertain. In this region, the \texttt{KST24} model stands out as it is the first to fit the Fan region rather than masking it out. Fitting this region necessitates a stronger magnetic field in the direction of the outer Galaxy compared to previous models, resulting in significantly larger deflections in this area. This in turn can affect reconstruction of the UHECR dipole.

In summary, in most directions on the sky, UHECR deflections are still too uncertain --- both in direction and magnitude --- to enable source identification through backtracking. A notable exception is the region at high latitudes, roughly at $30^\circ < b <  70^\circ$ in the outer Galaxy, where all existing models consistently predict relatively small deflections by the coherent GMF. Interestingly, this is the same region where the TA hotspot has been observed. Regarding the Amaterasu particle, the precise localization of its source cannot be achieved with the \texttt{KST24} model, although the reconstructed source region is close to that found with the \texttt{UF23} model.

Most of the uncertainty in current GMF modeling, at least within the phenomenological approach discussed here, arises not from the statistical limitations of the data but from its interpretation and freedom to select different components and their parametrizations. This flexibility is, in turn, linked to the integral nature of the data used, where distance information is implicit. The resulting degeneracy could be resolved by adding data that explicitly include distance information, such as Faraday rotation measures of pulsars. Incorporating such data into future models should enable more precise predictions of UHECR deflections.

\vskip 0.cm

\section*{Acknowledgments}
\noindent We are grateful to Michael Unger for useful discussions. The work of DS has been supported in part by the French National Research Agency (ANR) grant ANR-24-CE31-4686. The work of AK and PT is supported by the IISN project No. 4.4501.18.

\bibliographystyle{apsrev4-2}
\bibliography{ref_GMF_UHECR}

\end{document}